\documentclass{article} 
\usepackage{iclr2023_conference,times}
\usepackage{graphicx}
\usepackage{enumitem}
\usepackage{placeins}

\usepackage{color}
\usepackage{array}
\usepackage{tabu}
\usepackage{vcell}

\usepackage{tabu}
\usepackage{amsmath}

\usepackage{amsmath,amsfonts,bm}

\def\eqref#1{equation~\ref{#1}}

\def\1{\bm{1}}

\DeclareMathAlphabet{\mathsfit}{\encodingdefault}{\sfdefault}{m}{sl}
\SetMathAlphabet{\mathsfit}{bold}{\encodingdefault}{\sfdefault}{bx}{n}

\usepackage[hyphens]{url}
\usepackage[hidelinks]{hyperref}
\hypersetup{breaklinks=true}
\title{Uncovering the Spatial and Temporal Variability of Wind Resources in Europe: A Web-Based Data-Mining Tool}

\author{Alban Puech \\
Institut Polytechnique de Paris\\
Palaiseau, France \\
\texttt{alban.puech@polytechnique.edu} \\
\And
Jesse Read \\
Laboratoire d'informatique de l'École Polytechnique\\
Palaiseau, France \\
\texttt{jesse.read@lix.polytechnique.edu} \\
}

\makeatletter
\newcommand\newtag[2]{#1\def\@currentlabel{\textcolor{blue}{#1}}\label{#2}}
\makeatother
\iclrfinalcopy 
\begin{document}
\noindent
\maketitle

\begin{abstract}
We introduce \href{https://www.remap-eu.app/}{\textcolor{blue}{REmap-eu.app}}, a web-based data-mining visualization tool of the spatial and temporal variability of wind resources. It uses the latest open-access dataset of the daily wind capacity factor in 28 European countries between 1979 and 2019 and proposes several user-configurable visualizations of the temporal and spatial variations of the wind power capacity factor. The platform allows for a deep analysis of the distribution, the cross-country correlation, and the drivers of low wind power events. It offers an easy-to-use interface that makes it suitable for the needs of researchers and stakeholders. The tool is expected to be useful in identifying areas of high wind potential and possible challenges that may impact the large-scale deployment of wind turbines in Europe. Particular importance is given to the visualization of low wind power events and to the potential of cross-border cooperations in mitigating the variability of wind in the context of increasing reliance on weather-sensitive renewable energy sources. 
\end{abstract}
   
\section{Introduction}
As the world is facing an unprecedented energy crisis, more and more investments are going into intermittent renewable energy sources (IRES). \citeauthor{irena}'s \citeyearpar{irena} analysis forecasts wind power to represent 40 percent of the EU-28 electricity needs in 2050. As energy production depends more and more on weather conditions, there is a call for tools that could help in better understanding and dealing with the risks that this can present. 
Wind variability poses a significant challenge to energy production, as it can lead to fluctuations in power output, making it difficult to manage energy supply \cite{impact}. Therefore, understanding the historical variability of wind power is essential to make well-informed decisions regarding the future deployment and management of wind turbines. One solution to the problem of wind variability is cross-border cooperation, which allows countries to share their wind resources and balance out variations in power output. In January 2023, the European Commission called for cross-border renewable energy projects \cite{EU}, and more and more attention is given to the study of the existing and the future necessary regulations, policies, and incentives \cite{regulation,publication_2023,eeagency_2020,policy_brief}. However, to effectively implement cross-border cooperation, a deep understanding of the spatial and temporal distribution of wind energy in different countries and regions is needed. This is where data-mining visualization tools can play a crucial role by providing an intuitive and interactive way to explore and analyze wind data from different regions, helping to identify patterns, trends, and potential issues that would be difficult to discern from raw data alone. 
 
In this paper, we present a dashboard-like data-mining interface that aims at making the historical spatial and temporal variability of wind power more accessible to decision-makers. We also hope to bridge the gap between the available climate data and the needs of the energy industry looking for a more intelligible analysis tool that: (1) Uses the latest open-access dataset of the daily wind capacity factor in 28 European countries between 1979 and 2019; (2) Proposes more than 16 fully-configurable visualizations of the temporal (intraday, intrayear, year-over-year), and spatial variations of the wind power capacity factor; (3) Allows a deep analysis of the distribution, the cross-country correlations of the temporal distribution, and the drivers of low wind power events of different European countries.


We expect the platform to be useful to (1): Researchers seeking visual representations of the historical variability of wind power, the possible obstacles to the reliability of a power supply system highly relying on wind, and more generally to the large-scale integration of wind power into the grid; provide stakeholders (2) with valuable insights into the potential of countries for wind energy development, as well as the potential for regional cooperation and help them make well-informed decisions.


\section{Dataset, development and deployment specifics}
\textbf{Dataset.} The dataset used here is an ERA5-derived time series of European country-aggregate electricity wind power generation \cite{dataset}. The authors use the ERA5 reanalysis data \cite{ERA5} to compute the nationally aggregated hourly capacity factor of 28 European countries between 1979 and 2019. The capacity factor of a wind turbine is the ratio of the electrical energy produced by the turbine for the period of time considered to the total amount of electrical energy that could have been produced at full power operation during the same period. Further details regarding the data are given in appendix A1. 

\textbf{Software Framework.} Our platform is built using Dash, a python library built on top of React.js that allows building data applications with perfect integration of Plotly.js. Dash was made popular for its use in bioinformatics database visualization \cite{dash}. Dash uses callbacks to modify or add new HTML elements as a result of the user's action. The platform is accessible at: \href{https://www.remap-eu.app/}{\textcolor{blue}{REmap-eu.app}}

\textbf{Interactive Two-Card-Layout. }
The layout of the web application is based on two cards, placed side-by-side. Fig. \ref{fig:cards} in the Appendix shows a screenshot of the platform. Each card groups together elements that interact with each other. Hence, each of the two cards is independent, with a unique exception: The choropleth map of the left card is used to filter the countries plotted on the right card. The left card displays `raw' data, with little pre-processing involved. It provides a simple yet clear description of the spatial and temporal distribution of the capacity factor. The right card provides more elaborated visualizations, that require more pre-processing. Those visualizations are to be interpreted with the data shown on the left card. The layout was designed to allow the two plots to be side by side, so that the user does not need to switch between them.

\section{Visualizations for the spatial and temporal analysis of wind power}
\label{section3}
In this section, we introduce the different visualizations describing the spatial and temporal variability of wind energy. Appendix A2 gives a summary of the proposed plots and their settings.

\textbf{Analysing the spatial distribution of wind resources in Europe.}
European countries are not equal in terms of wind resources \cite{potential}. We use a choropleth map \ref{a:chor} of the country-aggregated capacity factors to highlight the spatial distribution of wind power. We aim at providing an easy way to compare the average and the standard deviation of the capacity factor of European countries, for different time resolutions and over different time periods. The map shows the average and standard deviation of the capacity factor of each European country. This type of visualization is often used to show how different geographic entities compare. The settings available to the user are summarized in Appendix A2 and Fig. \ref{fig:choropleth} shows a possible configuration of such a map.

\textbf{Comparing the temporal distribution of wind resources across countries.} We display a line plot of the capacity factor \ref{a:line1} below the choropleth map. The role of the line plot is to show the intraday, intrayear, and yearly capacity factor variations of European countries, depending on what resolution the user chose. It also allows comparison of the capacity factor across countries, over different time windows. The average capacity factor over all 28 European countries available is displayed, and the line is labeled as ``28C". This allows the user to compare a country with the Europe-aggregated data, or compare multiple countries, as shown in Fig. \ref{fig:lineplot}.

\begin{figure}[!htb]
    \centering
\includegraphics[width=0.5\linewidth]{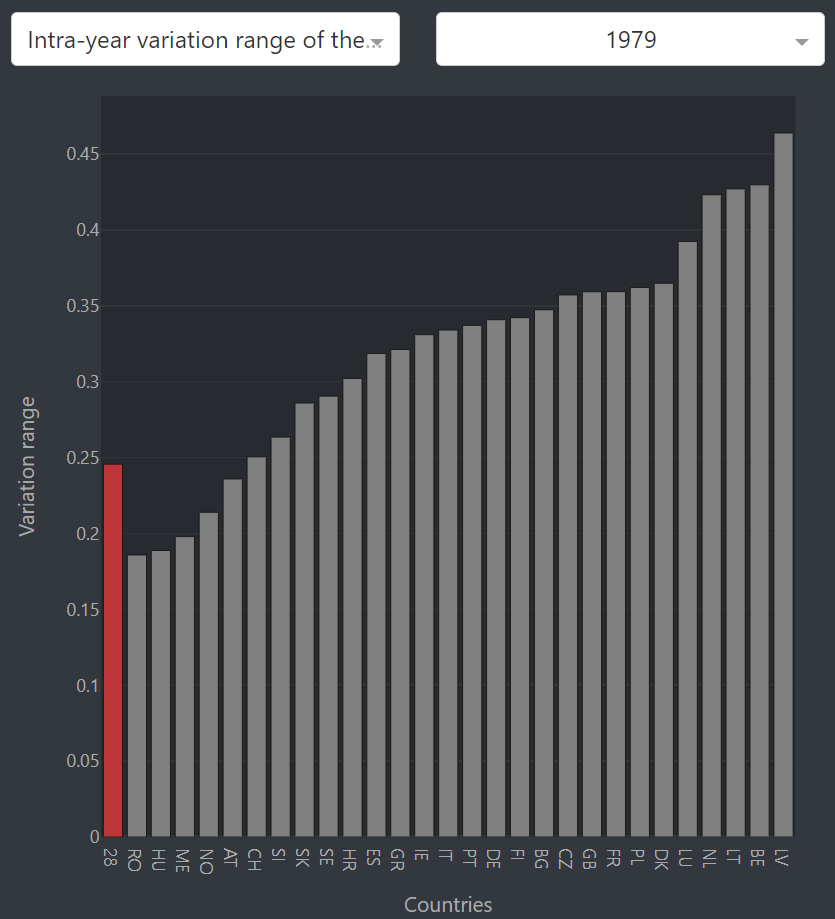}
  \caption{Bar chart of the intrayear variation range of the monthly capacity factor of the 28 countries in 1979. The bar corresponding to the 28 countries-aggregated capacity factor range is shown in red. This chart helps in identifying countries that suffer from a large capacity factor gap between high and low wind power months.}
  \label{fig:intrayear}
\end{figure}

\textbf{Intrayear and intraday wind resources variability.} The ``Intrayear variation range of the monthly capacity factor" bar plot \ref{a:variability} displays the variation range of the monthly capacity factor of each country, as shown in Fig. \ref{fig:intrayear}. The same plot is provided for the intraday variation range of the hourly capacity factor. Understanding seasonal patterns in historical data can help in building more effective forecast models, which in turn allows for more accurate planning and management of the grid. This knowledge can also be used to develop more effective policies and strategies for integrating wind energy into the grid \cite{balancing} and maximizing its potential benefits. For example, if hybrid energy systems based on solar and wind are often assumed to better deal with the variability of renewable energies, \citet{summer_winter} showed that in some countries, solar and winter may not complement each other as well. This is the case when wind production does not increase during the winter, and as a consequence, does not compensate for the drop in solar energy production. 

\begin{figure}[!htb]
    \centering
  \includegraphics[width=.75\linewidth]{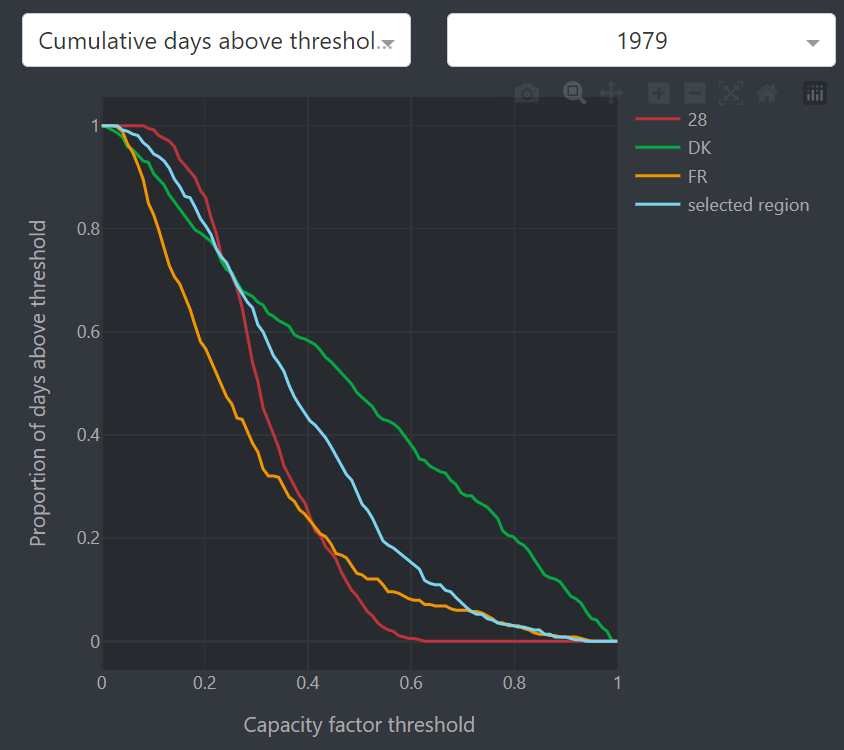}
  \caption{Line plot of the cumulative days above thresholds for the year 1979. We display the data for France and Denmark, and the data aggregated over the region (France+Denmark) in blue. When neighboring countries are connected, excess energy from windy regions can be transferred to regions with less wind, reducing the need for reserve capacity. In practice, this can lead to an increase in the overall capacity factor of wind power in the interconnected system.}
  \label{fig:cum}
\end{figure}

\textbf{Cumulative time above threshold comparison.} One way to compare the capacity factor of different regions is to look at their proportion of days that had a capacity factor higher than some threshold. This is what can be done using the ``Cumulative days above threshold" plot \ref{a:cumulative}. The plot supports multi-country selection, in which case a line corresponding to the data aggregated over the entire selected region is added, as in Fig. \ref{fig:cum}. Again, this allows the user to consider the selected countries as a single unique region in the context of perfect grid interconnection. 

\begin{figure}
    \centering
  \includegraphics[width=.75\linewidth]{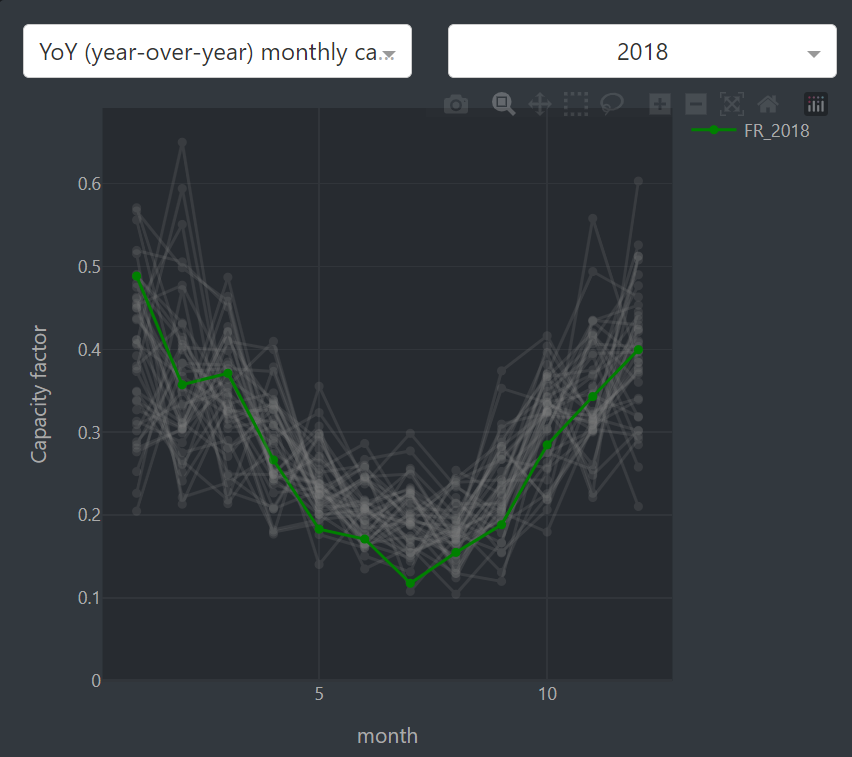}
  \caption{Year-over-Year monthly capacity factor comparison for France, The highlighted year is 2018. The gray lines correspond to the other years of the period (1979-2019). 2018 showed a particularly low wind production during summer (note how the height of the green line compares to the gray lines). }
  \label{fig:yoy}
\end{figure}

\textbf{Year-over-year monthly capacity factor comparison.} So far, the visualizations were focusing on the spatial distribution of wind power, and on the comparison of capacity factor-derived features across countries. However, the increasing investments into wind energy have pushed for more research on the year-over-year country-scale evolution of wind energy resources \cite{yoy}. The long-term evolution of wind resources is important for the calculations used in the preliminary assessment of energy-producing projects, such as the levelized cost of energy (LCOE). In the ``YoY (year-over-year) monthly capacity factor comparison" plot \ref{a:yoy}, we display the intra-year evolution of the capacity factor for the selected country and the selected year.  The lines corresponding to the other years of the period 1979-2019 are displayed in gray, allowing the user to compare the capacity factor of a given year to the other years of the period. This is shown in Fig. \ref{fig:yoy}.  

\section{Analyzing low wind power events: Frequency, drivers and mitigation strategies}

Unlike solar PV, which exhibits relatively predictable diurnal and seasonal cycles, wind power has more complex and irregular variations in energy generation, both at inter-annual and intra-annual scales. In particular, the study of the temporal distribution of low-wind-power (LWP) events has gained more attention in the literature \cite{intermittency, lwpDE}. LWP events are becoming a growing concern in countries where wind power makes up a significant portion of the energy mix, as it raises questions about energy security and stability. Studying the past occurrences of these events can provide valuable insights into the drivers of variability and inform the development of strategies to mitigate their impact. Although there is not a single definition of low wind power events, they can be defined as an uninterrupted period of time during which the capacity factor is below some threshold \cite{LWPdef1,LWPdef2,lwpDE}. In our web app, we arbitrarily set this threshold to 10 percent. This is a value that will be modifiable by the users in the next version of the platform.

\begin{figure}
    \centering
  \includegraphics[width=.75\linewidth]{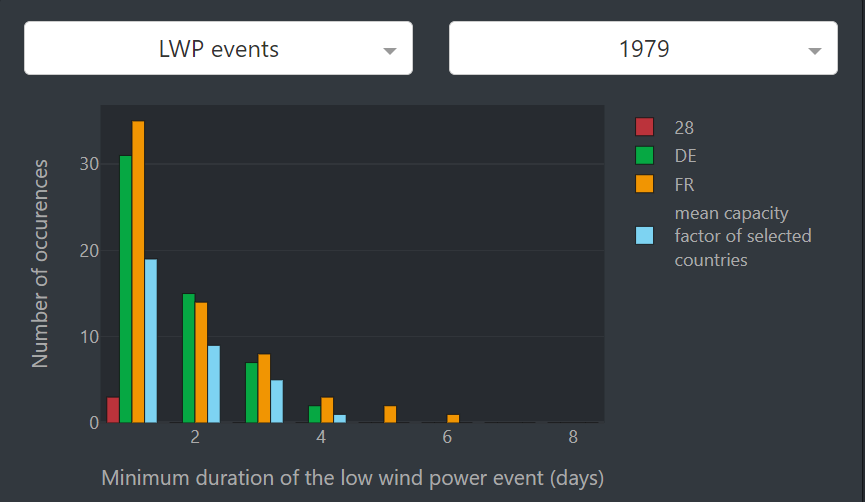}
  \caption{Number of low wind power events in France, Germany, and in the entire (France+Germany) region for different minimum LWP events duration in 1979. Interconnected countries have a tendency to experience fewer low wind power events due to the ability of one country to compensate for the other's low capacity factor, particularly when they have distinct coastal regions and are exposed to varying wind patterns. Grid interconnections generally enhance the reliability of energy supply.}
  \label{fig:lwp}
\end{figure}

\begin{figure}
    \centering
  \includegraphics[width=.75\linewidth]{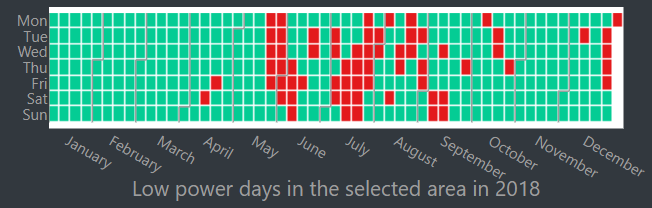}
  \includegraphics[width=.75\linewidth]{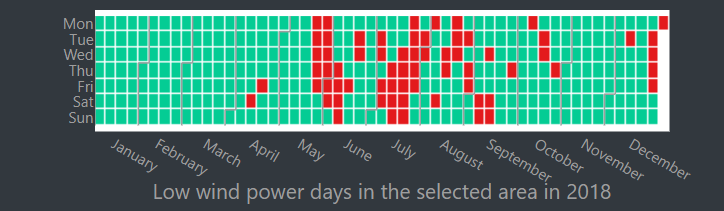}
  \caption{Calendar plots of the low wind power days in France (top figure), and in the region (France+Germany) in 2018. Low wind power days are displayed in red. The observation is similar to the one made about Fig. \ref{fig:lwp}.}
  \label{fig:cal}
\end{figure}

\textbf{Comparing the number, the duration, and the distribution of low wind power events.}
When the user selects ``LP events" in the dropdown, two plots are displayed. The first one is a bar plot of the number of occurrences of low wind power events for different minimum durations. The second one is a calendar plot that indicates the low wind power days in the selected region. When the user selects multiple countries, the bar chart displays grouped bars corresponding to each country. This allows for comparing the number of occurrences of LWP events of each minimum duration across selected countries. We also add the data corresponding to the selected-region-aggregated data. This allows the user to see how grid interconnection mitigates the risk of observing LWP events, as shown in Fig. \ref{fig:lwp}. Indeed, a selected region often has a lower number of LWP events than each of its constitutive countries, since LWP events don't necessarily happen at the same time in all constitutive countries. The calendar plot indicates the low wind power days at the scale of the selected region. This plot gives information on the temporal distribution of those days within the considered year. An example of such plots is shown in Fig. \ref{fig:cal}.

\begin{figure}
    \centering
  \includegraphics[width=.75\linewidth]{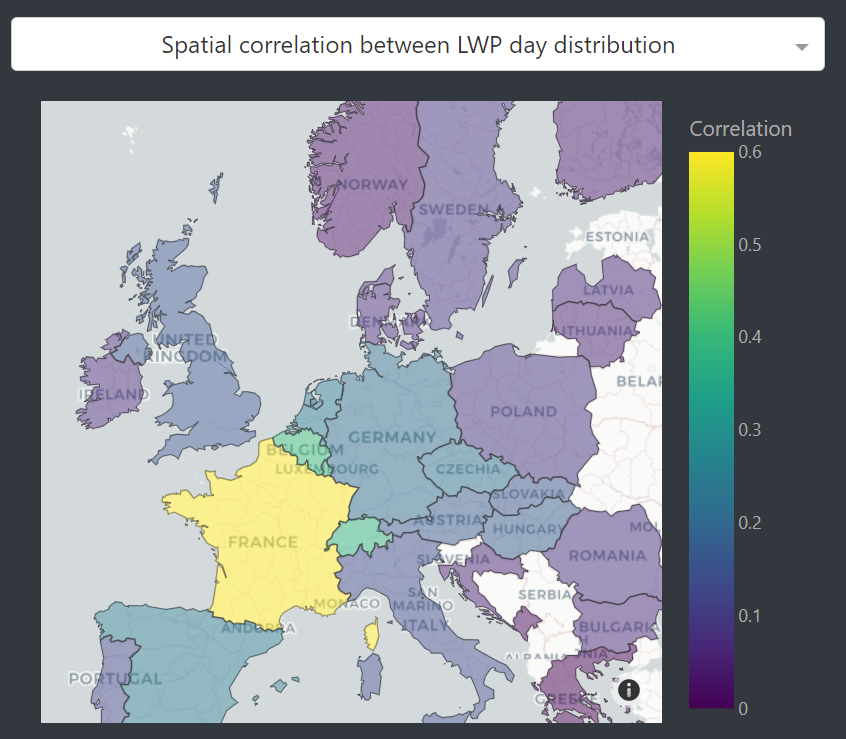}
  \caption{Correlation between the LWP events distribution of France and the ones of the 28 other European countries.}
  \label{fig:corr}
\end{figure}

\textbf{Cross-country correlation of the low wind power day temporal distribution.} 
The previous plots that we described allow the user to compare countries in terms of their number of LWP events. However, understanding how the capacity factors of neighboring countries are correlated is of major importance to determine the interconnection level that could help in alleviating the spatial variability of wind energy \cite{corr1,corr2,corr3}. For example, by understanding the patterns and trends of low wind power events in different countries, it may be possible to identify the most appropriate market mechanisms, such as interconnection capacities, pricing schemes, and balancing mechanisms, that can enable cross-border cooperation. For this reason, we propose two different choropleth maps. Both of them require one country to be selected on the left card. The first plot shows the Pearson correlation coefficient between the selected country and the other countries in terms of LWP day distribution, see Fig.\ref{fig:corr}. The second one shows the same statistics, but for the raw capacity factor values. Only statistically significant (p-value$\ge$0.05) correlations are displayed.

\begin{figure}
    \centering
  \includegraphics[width=.75\linewidth]{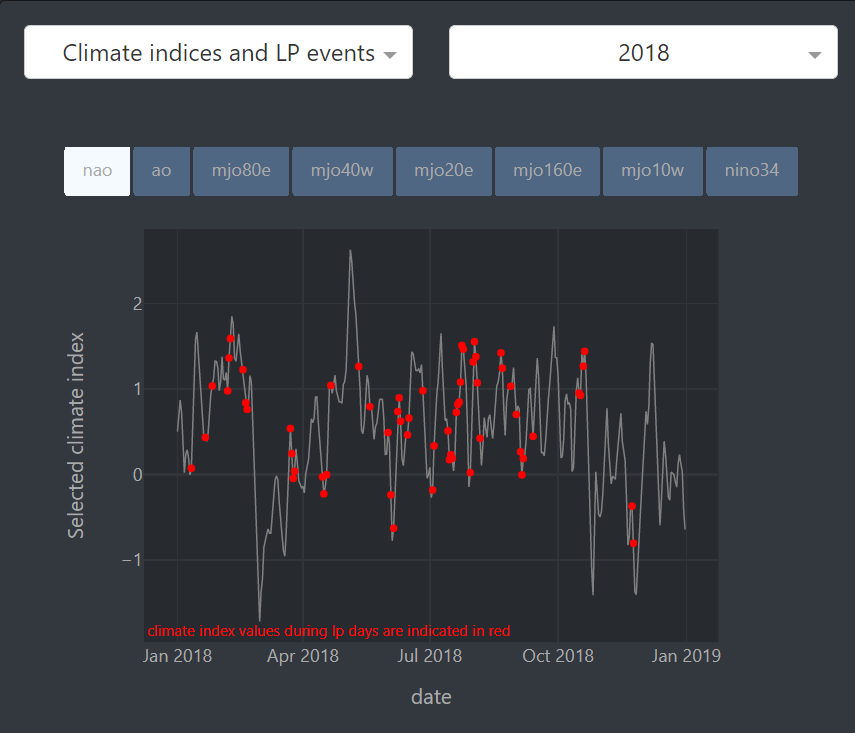}
  \caption{Line plot of the NAO climate index in 2018. The red dots correspond to low wind power days in the selected country (here, France). After having identified a low wind power event using the calendar plot shown in Fig. \ref{fig:cal}, a typical workflow would be to use this visualization to investigate the climate indices preceding this event.}
  \label{fig:indices}
\end{figure}

\textbf{Detecting the possible drivers of low wind power events. }
Climate indices provide a measure of large-scale atmospheric circulation and weather conditions that impact wind resources. For instance, the North Atlantic Oscillation (NAO) is one of the most commonly studied climate indices that has a positive correlation with wind power in Europe \cite{nao1,nao2}. We incorporate a plot of the historical climate indices during low wind speed events, providing a valuable tool for examining the relationship between climate indices and wind generation. It can help to identify the indices that are most correlated with wind generation. This historical information can be used to develop more accurate models for predicting wind power generation. The plot shows the climate indices for the selected year and highlights the values corresponding to low wind power days in the selected country, as shown in Fig. \ref{fig:indices}. The user can select the climate index to display among the North Atlantic Oscillation index (NAO), the Artic Oscillation index (AO), the Madden-Julian Oscillation indices (MJO) for different longitudes, and the El-Niño Southern Oscillation (NINO). 

\section{Additional features}
\begin{figure}
    \centering
  \includegraphics[width=.75\linewidth]{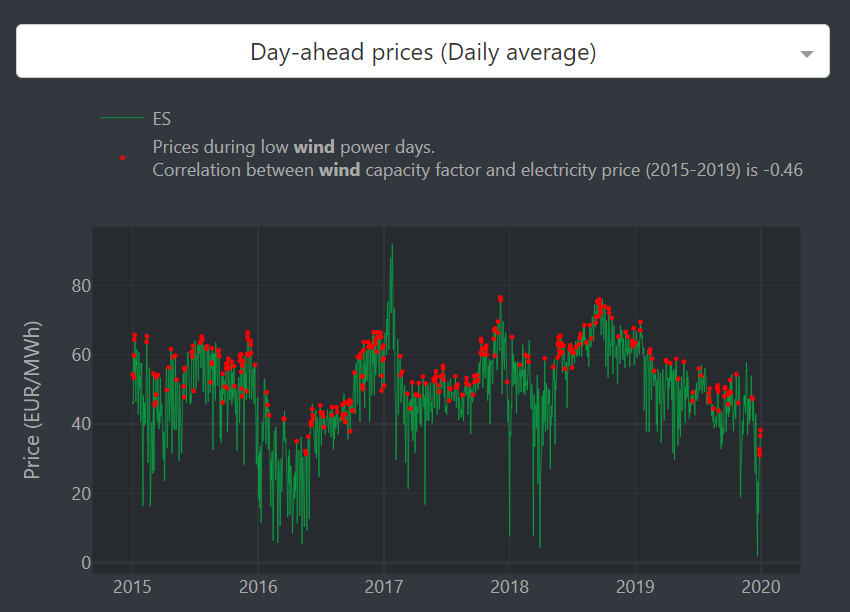}
  \caption{Line plot of the electricity prices in Germany. The red dots correspond to low wind power days. The correlation between wind capacity factor and electricity prices is particularly high in countries that highly rely on wind, as is the case of Germany. }
  \label{fig:prices}
\end{figure}

\textbf{Electricity prices.}  It is important to consider the interplay between wind power, electricity prices, and other factors in order to develop a comprehensive understanding of the energy market. Low wind power days can have a significant impact on electricity prices. When wind power generation decreases, other sources of electricity, such as fossil fuels or hydropower, need to ramp up production to compensate. This can result in an increase in electricity prices, as those energy sources typically have a variable cost, contrary to wind and solar which have no fuel or variable O\&M costs. We propose a plot of the daily average day-ahead electricity prices, shown in Fig \ref{fig:prices}. We highlight the prices corresponding to low wind power days and display the correlation between low wind power events and electricity prices, which is found to be high for European countries that heavily rely on this energy source. The price data is obtained from \citet{price}.

\textbf{Solar energy data.} Combining wind and solar energy has gained interest as a way to mitigate their intermittency and variability, creating a more reliable and stable energy mix. The platform proposes the user to compare different renewable energy mixes by choosing the weights given to solar and wind energy in the computation of the capacity factor data displayed in the visualizations. The solar capacity factor data is also obtained from \citet{dataset}. By default, only the data corresponding to wind energy (respective weights of 1 and 0 for wind and solar) is displayed, and we only focused on the visualizations obtained using this setting in this paper.

\section{Conclusion and Future Work}
In this paper, we presented a new web platform that offers multiple visualizations of the temporal and spatial variability of historical wind energy resources over Europe. The tool proposes configurable plots that allow the user to deeply analyze the ERA5-derived capacity factor dataset \cite{dataset}. Although we primarily expect this platform to be useful to climate researchers, the energy industry, and the decision-makers, we also hope to serve the needs of machine learning engineers and scientists looking for a better understanding of the wind energy resource assessment challenges.
We plan on continuously improving the platform based on the feedback that we have already received from academics and stakeholders who were introduced to the tool. Specifically, we will add the demand data. This will allow the study of the relationship between electricity demand and wind production. We also plan on adding the demand-net-wind, the electricity demand that needs to be covered by another energy source than wind.

\section{Acknowledgments} We thank Dr. Naveen Goutham for his help in correcting the last version of the paper. We also thank Dr. Hannah Bloomfield and Prof. Emmanuel Pietriga for their insights and suggestions to improve the platform.
\bibliography{iclr2023_conference}
\bibliographystyle{iclr2023_conference}
\clearpage
\appendix
\section{Appendix}

\subsection{A1 - Dataset}
The authors of the dataset \cite{dataset} used by the platform use the ERA5 reanalysis data \cite{ERA5} to compute the nationally aggregated hourly capacity factor of 28 European countries between 1979 and 2019.
Reanalysis combines past weather observations and current atmospheric models to generate climate and weather historic data \cite{ReanalysesandObservationsWhatstheDifference}. It allows getting a complete weather record from sparse - both in space and time - past data. In addition to the wind speed data, the authors use the wind farm spatial distribution of 2017, taken from \url{https://thewindpower.net}.
However, it is worth mentioning that, because the absolute wind power capacity is not used to compute the capacity factors, only the relative spatial distribution of wind turbines is assumed to be constant. The capacity factor of each country is estimated by aggregating the capacity factor computed for each grid box, weighted by its estimated installed capacity. The capacity factor in each grid box is derived using the 100 m wind speeds and the power curve of the type of wind turbine maximizing the energy produced during the entire period (1979-2019), as indicated in \cite{forecast}. Although the authors reported an average percentage error of 10 percent in the validation settings, it is important to note that capacity factors indicated here may deviate from the true values.

\clearpage
\subsection{A2 - Summary of the visualizations described in Section \ref{section3}}
\FloatBarrier

\begin{table}[!htb]
\centering
\resizebox{!}{0.4\textheight}{%
\begin{tabular}{|>{\hspace{0pt}}p{0.217\linewidth}|>{\hspace{0pt}}p{0.235\linewidth}|>{\hspace{0pt}}p{0.49\linewidth}|} 
\hline
\multicolumn{1}{|>{\centering\hspace{0pt}}m{0.217\linewidth}|}{Description}                                                  & \multicolumn{1}{>{\centering\hspace{0pt}}m{0.235\linewidth}|}{User interaction}                                                                                                                                                                        & \multicolumn{1}{>{\centering\arraybackslash\hspace{0pt}}m{0.49\linewidth}|}{Comments}                                                                                                                                                                                                                                                                                                                                                                                                                                                                                        \\ 
\hline
\vcell{\newtag{\textcolor{blue}{[1]}}{a:chor} Choropleth map of~the average or standard deviation of the hourly,~~monthly or yearly capacity factor } & \vcell{- The right dropdown sets the data resolution (yearly, monthly, or hourly).\par{}- The left dropdown allows switching between average and std of the capacity factor\par{}- The slider filters the year, month or hours included in the data}   & \vcell{- The Average and standard deviation are computed over the entire time period~(1979-2019), except if the time resolution is set to yearly, in which case the user can decide what years to include}                                                                                                                                                                                                                                                                                                                                                                   \\[-\rowheight]
\printcelltop                                                                                                                & \printcelltop                                                                                                                                                                                                                                          & \printcelltop                                                                                                                                                                                                                                                                                                                                                                                                                                                                                                                                                                \\ 
\hline
\vcell{\newtag{\textcolor{blue}{[2]}}{a:line1} Line plot of the intraday, intrayear or yearly average capacity factor }                               & \vcell{- The countries displayed on the plot are those selected when clicking on the choropleth map (holding shift allows multi-selection)\par{}~- The range slider can then be used to filter the time period that is displayed}                      & \vcell{- When no country is selected (initial state), the average capacity factor over all 28 European countries available is displayed, and the line is labelled as ``28C"}                                                                                                                                                                                                                                                                                                                                                                                                  \\[-\rowheight]
\printcelltop                                                                                                                & \printcelltop                                                                                                                                                                                                                                          & \printcelltop                                                                                                                                                                                                                                                                                                                                                                                                                                                                                                                                                                \\ 
\hline
\vcell{\newtag{\textcolor{blue}{[3]}}{a:variability} Bar plots of the Intrayear/intraday variation range of the monthly/ hourly capacity factor}     & \vcell{- The dropdown located on the top part of the card is used to select the year of the data to be shown\par{}- The countries displayed on the plot are those selected when clicking on the choropleth map (holding shift allows multi-selection)} & \vcell{- When no country is selected on the left card, all European countries are plotted, in ascending order of values. This is because the goal of this visualization is to provide a ranking of the countries based on their capacity factor variability\par{}- The scale starts at 0 for the difference in bar heights to accurately represent the difference in capacity factor variation range.\par{}- The color used for each country is consistent across all visualizations of the web app, allowing the user to quickly spot the selected countries on each plot}  \\[-\rowheight]
\printcelltop                                                                                                                & \printcelltop                                                                                                                                                                                                                                          & \printcelltop                                                                                                                                                                                                                                                                                                                                                                                                                                                                                                                                                                \\ 
\hline
\vcell{ \newtag{\textcolor{blue}{[4]}}{a:intra} Plot of the minimum, maximum and average intraday/ intrayear capacity factor~}                                        & \vcell{as in 3.}                                                                                                                                                                                                                                       & \vcell{- The scatter plot displays the mean with a cross and the min and max values with error bars\par{}- The countries are ordered by mean values}                                                                                                                                                                                                                                                                                                                                                                                                                         \\[-\rowheight]
\printcelltop                                                                                                                & \printcelltop                                                                                                                                                                                                                                          & \printcelltop                                                                                                                                                                                                                                                                                                                                                                                                                                                                                                                                                                \\ 
\hline
\vcell{\newtag{\textcolor{blue}{[5]}}{a:cumulative} Line plot of the cumulative days above capacity factor threshold}                           & \vcell{as in 3. Moreover :\par{}- When multiple countries are selected, an additional line corresponding to the selected region (average over all selected countries) is displayed}                                                                    & \vcell{- When no country is selected, the data for the 28 countries-aggregated capacity factor is displayed, and the lines corresponding to each European country are displayed in light gray.~This allows the user to quickly see how each country compares to the other 27 European countries by hovering over its corresponding line}                                                                                                                                                                                                                                     \\[-\rowheight]
\printcelltop                                                                                                                & \printcelltop                                                                                                                                                                                                                                          & \printcelltop                                                                                                                                                                                                                                                                                                                                                                                                                                                                                                                                                                \\ 
\hline
\vcell{\newtag{\textcolor{blue}{[6]}}{a:yoy} YoY (year-over-year) monthly capacity factor plot }                                                      & \vcell{as in 4.}                                                                                                                                                                                                                                       & \vcell{-~As the goal here is really to compare the intrayear capacity factor over the years for a single region, selecting multiple countries results in showing only one single line corresponding to the data aggregated over the entire selected region\par{}- The lines corresponding to the other years of the period 1979-2019 are displayed in light gray, allowing the user to quickly figure out how the capacity factor of a given year compares to the other years of the period}                                                                                 \\[-\rowheight]
\printcelltop                                                                                                                & \printcelltop                                                                                                                                                                                                                                          & \printcelltop                                                                                                                                                                                                                                                                                                                                                                                                                                                                                                                                                                \\
\hline
\end{tabular}
}
\end{table}

\FloatBarrier
\clearpage
\subsection{A3 - figures and use-cases examples}
\begin{figure}[!htb]
  \includegraphics[width=\linewidth]{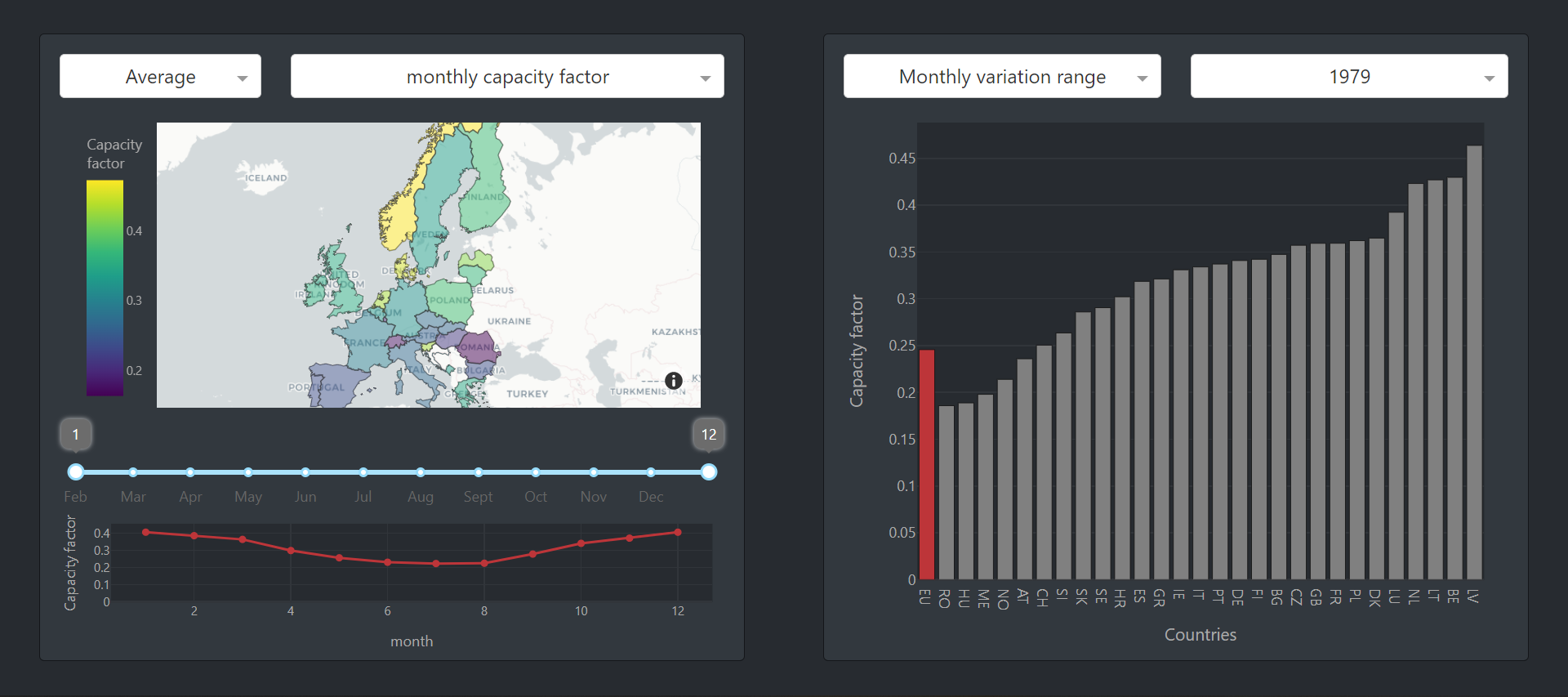}
  \caption{Layout of the web app. The left card shows the choropleth map \ref{a:chor} and the line plot \ref{a:line1} of the capacity factor, while the right one offers more advanced visualizations. The choropleth map of the left card is used to filter the data plotted on the right card.}
  \label{fig:cards}
  
\end{figure}
\begin{figure}[!htb]
\centering
  \includegraphics[width=0.65\linewidth]{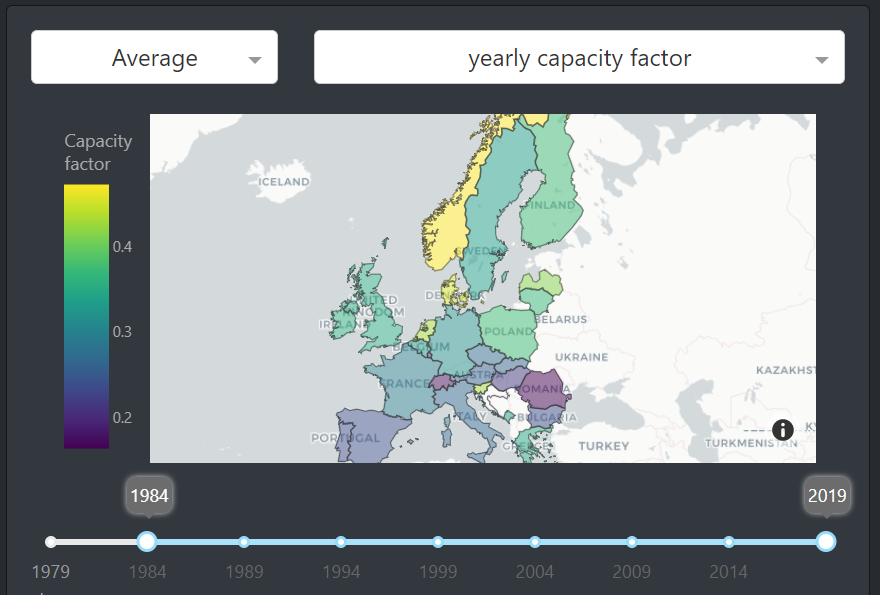}
  \caption{Choropleth map \ref{a:chor} of the average yearly capacity factor over the period 1984-2019. The range slider and the two dropdowns are used to parametrize the visualization.}
  \label{fig:choropleth}
\end{figure}

\begin{figure}[!htb]
\centering  \includegraphics[width=0.65\linewidth]{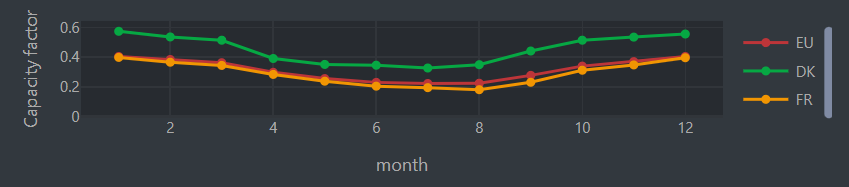}
  \caption{Line plot \ref{a:line1} showing the average monthly capacity factor of France, the UK, and the average capacity factor over the 28 European countries (28C) for the period 1979-2019. The scale starts at 0 to allow a better comparison between countries. The user selects the countries to be displayed by clicking on the choropleth map \ref{a:chor} shown in Fig. \ref{fig:choropleth} and sets the time resolution using the dropdown shown in Fig. \ref{fig:cards}.}
  \label{fig:lineplot}
\end{figure}

\end{document}